# Atomic-scale observation of geometric frustration in a fluorine-intercalated infinite layer nickelate superlattice

Chao Yang[1]*, Roberto A. Ortiz[1], Hongguang Wang[1], Wilfried Sigle[1], Kelvin Anggara[1], Eva Benckiser[1], Bernhard Keimer[1], Peter A. van Aken[1]

[1]Max Planck Institute for Solid State Research, Stuttgart, 70569, Germany
*Corresponding author: c.yang@fkf.mpg.de

**Abstract**

Anion doping offers immense potential for tailoring material properties, but achieving precise control over anion incorporation remains a challenge due to complex synthesis processes and limitations in local dopant detection. Here, we investigate the F-ion intercalation within an infinite layer $NdNiO_{2+x}$/$SrTiO_3$ superlattice film using a two-step synthesis approach. We employ advanced four-dimensional scanning transmission electron microscopy (4D-STEM) coupled with electron energy loss spectroscopy to map the F distribution and its impact on the atomic and electronic structure. Our observations reveal a striking geometric reconstruction of the infinite layer structure upon fluorination, resulting in a more distorted orthorhombic phase compared to the pristine perovskite. Notably, F-ion intercalation occurs primarily at the apical sites of the polyhedron, with some occupation of basal sites in localized regions. This process leads to the formation of two distinct domains within the nickelate layer, reflecting a competition between polyhedral distortion and geometric frustration-induced neodymium (Nd) displacement near domain interfaces. Interestingly, we observe an anomalous structural distortion where basal site anions are displaced in the same direction as Nd atoms, potentially linked to the partial basal site F-ion occupation. This coexistence of diverse structural distortions signifies a locally disordered F-ion distribution within the infinite layer structure with distinct F-ion configurations. These findings provide crucial insights into understanding and manipulating anion doping at the atomic level, paving the way for the development of novel materials with precisely controlled functionalities.

**Introduction**

Perovskite oxides have emerged as a versatile platform for exploring a multitude of fascinating phenomena by unlocking the intricate interplay between the crystal structure, electronic configuration, and processing conditions. These include colossal magnetoresistance(*1*), ferroelectricity and multiferroicity(*2*), and superconductivity(*3*). A powerful strategy to manipulate the properties of these materials lies in topochemical anion intercalation/exchange, where guest ions are introduced into the lattice while preserving the overall crystallographic framework. One particularly intriguing guest ion is fluorine (F), whose incorporation can dramatically alter material behavior. For instance, F-ion exchange/intercalation in $NdNiO_3$ thin films induces a transformation from a metallic to a highly insulating state, partially related to enhanced Coulomb repulsion in the Ni $3d$ orbitals, which is reversible after annealing in an oxygen atmosphere(*1*). Furthermore, first-principles calculations show that these fluorinated films exhibit a stronger orthorhombic distortion compared to the pristine $NdNiO_3$ perovskite(*4*). The calculations predict an ordered, anisotropic arrangement of F ions within the lattice(*4*), a feature crucial for understanding the unique properties of mixed anion compounds. Moreover, F-ion intercalation can lead to a rich variety of phenomena. For instance, Ruddlesden-Popper nickelate oxides ($n = 2$) exhibit a breaking of local inversion symmetry due to F-ion insertion, inducing a local antiferroelectric state(*2*). Similarly, the incorporation of F plays a crucial role in achieving superconductivity in $Sr_2CuO_2F_{2+x}$ cuprates(*5-8*). However, a critical gap in our understanding remains: the precise influence of F-ion distribution on the atomic structure of perovskite oxyfluorides and the difference between the O and F positions has not been directly observed experimentally.

Beyond anion intercalation, deintercalation via topochemical reactions offers another powerful strategy to manipulate the properties of perovskites. This approach modifies the atomic and electronic structures of the host material by removing specific anions. A prime example lies in the synthesis of infinite layer structures from ferrites(*9, 10*) and nickelates(*11-14*). Here, reducing agents like $CaH_2$ are employed to remove apical oxygen from the octahedra in their corresponding perovskite phases(*15*). The resulting infinite layer structures have garnered significant interest, particularly due to the recent discovery of superconductivity in $Nd_{0.8}Sr_{0.2}NiO_2$ films(*3*). This has spurred intense research to elucidate the similarities and differences between nickelates and cuprates in terms of magnetic structure(*16, 17*), superconducting behavior(*18, 19*), and charge ordering(*20-22*). Understanding these aspects is crucial for advancing the theory of high-temperature superconductivity. Notably, resonant X-ray scattering experiments revealed a commensurate charge density wave order at a wave vector of (0.333, 0) reciprocal lattice units in the infinite layer structure, suggesting a distinct multiorbital character compared to cuprates(*20*). However, a combination of X-ray scattering and atomic-scale observations has challenged this interpretation, suggesting the $3a_0$ superlattice peak might originate from an intermediate nickelate phase with residual apical oxygen ordering(*23*). This underscores the critical role of precise control over synthesis and the need for atomic-scale interrogation of local atomic and electronic structures to fully comprehend the physical properties of these complex systems.

The intricate nature of topochemical reactions necessitates the exploration of alternative strategies for achieving precise control over anion ordering in oxyfluorides. A promising approach may lie in the combination of oxygen deintercalation and subsequent fluorine intercalation. This sequential approach offers the potential to engineer specific oxygen/fluorine arrangements, paving the way for a controllable manipulation of functional properties in these materials. Notably, the partial substitution of F for O can induce an electronic configuration of Ni $3d^{8.8}$ in the nickelate system, which has been theoretically linked to superconductivity. However, to fully exploit the potential of F-ion intercalation, a comprehensive understanding of the O/F configuration, its influence on atomic-scale structural geometry, and the resulting electronic structure is crucial.

In this work, we address this critical knowledge gap by investigating F-ion intercalation in infinite layer nickelates within an $8NdNiO_2/4SrTiO_3$ superlattice structure, building upon our prior work on the synthesis of this material(*13*). To achieve atomic-level resolution of the oxygen and fluorine distribution, we employed advanced microscopy techniques – specifically, aberration-corrected scanning transmission electron microscopy (STEM) coupled with electron energy-loss spectroscopy (EELS). Furthermore, we utilized the integrated center of mass (iCoM) imaging technique to map the anion positions with sub-angstrom precision. This multifaceted approach revealed a fascinating interplay between polyhedral rotations and geometric frustration arising from Nd displacements induced by the F-ion incorporation. We will subsequently discuss the implications of the observed F-ion distribution on the structural distortions and electronic structure, drawing insights from density functional theory (DFT) calculations.

**Results and Discussion**

Figure 1a schematically illustrates the two-step synthesis employed to create a fluorine-intercalated infinite layer nickelate. The first step involves the formation of the infinite layer structure within the $8NdNiO_2/4SrTiO_3$ superlattice through a topochemical reduction process utilizing $CaH_2$ as a reducing agent. This step entails deintercalation, where apical oxygen is predominantly removed from the oxygen octahedra within the $NdNiO_3$ layer (detailed in Figure S1 and our previous work(*13*)). A recent study of a larger volume fraction of the sample by linear-dichroic resonant x-ray reflectometry, however, reveals an about 30% fraction of the nickelate, where the basal oxygen is removed instead of the apical one(*24*). Notably, the infinite layer structure achieves stability within the confined thickness of the superlattice or when employing a capping layer(*13*). Subsequently, F atoms readily intercalate into the vacant apical positions within the infinite layer structure, forming a $NiO_xF_y$ polyhedron.

Figure 1b presents a high-angle annular dark-field (HAADF) image of the atomic structure within the $NdNiO_2$ and $SrTiO_3$ layers of the superlattice. Compared to bulk $NdNiO_3$ (out-of-plane spacing of 3.86 Å), the $NdNiO_2$ layer exhibits a significant reduction in this spacing to around 3.2 Å. Additionally, the characteristic zigzag arrangement of A site cations is absent in the infinite layer structure. The corresponding maps and line profiles for the zigzag angle remain consistent across both the $SrTiO_3$ and $NdNiO_2$ layers, strongly suggesting successful deintercalation of the apical oxygen anions. In contrast, Figure 1c showcases the HAADF image of the structure after fluorine intercalation into the $NdNiO_2$ layer. This incorporation of fluorine induces substantial structural modifications, evident from the expansion of the out-of-plane lattice spacing and the reappearance of a zigzag cation arrangement. Notably, the corresponding zigzag angle map on the right of Figure 1c reveals an inhomogeneous distribution of the zigzag arrangement of Nd atoms across different F-intercalated $NdNiO_2$ layers.

Furthermore, the displacement of Nd atoms within the same F-intercalated $NdNiO_2$ layer exhibits inconsistencies. The zigzag angle profile on the right side of Figure 1c reveals that the Nd atoms in the inner layer reach a maximum zigzag angle of around 9°, signifying a substantial rotation of their octahedra. This rotation is slightly stronger compared to bulk $NdNiO_3$ (~8.5°)(*25*). Interestingly, the angle sharply decreases to about 2° at the interfaces with the neighboring layers, suggesting a return to octahedral coupling at these boundaries. Figure 1e highlights regions within the F-intercalated $NdNiO_2$ layer that exhibit an apparent zigzag arrangement of Nd atoms, absent in the reduced structure (Figure 1d). These observations, along with the broader overview provided in Figure S2, point towards the formation of domains within the $NdNiO_xF_y$ layers characterized by different zigzag arrangements of Nd atoms. This inhomogeneous distribution suggests that strong structural coupling might occur not only at the interfaces between layers but also within the domain wall regions themselves.

While conventional HAADF microscopy struggles with oxygen imaging, the latest 4D-STEM technique offers a powerful solution. By reconstructing iCoM images, we achieved high-resolution visualization of the anion sublattice, enabling precise determination of structural distortions(*26*). Figure 2a showcases the reconstructed ADF image from the 4D-STEM data, revealing the high-quality atomic structure of the $8NdNiO_xF_y/4SrTiO_3$ superlattice. The corresponding iCoM image in Figure 2b clearly depicts the anion sublattice, highlighted by the prominent zigzag arrangement of Nd atoms and a pronounced octahedral distortion within the regions delineated by yellow dashed boxes. To quantify this distortion, the zigzag angle map of Nd atoms, shown in Figure 2c, utilizes the Sr atoms' map as a reference. The averaged zigzag angle line profile in Figure 2d reveals significant variations, with the highest angle reaching ~12° and gradually decreasing to zero within the $SrTiO_3$ layer. Notably, the zigzag angle in the $NdNiO_xF_y$ layer significantly surpasses that observed in perovskite $NdNiO_3$, signifying a strong structural distortion induced by F-ion intercalation. We further quantified the in-plane B-O-B angle (B: Ni and Ti) in Figure 2e. Notably, a strong correlation emerges: regions with a lower B-O-B angle in the NdNiO$_x$F$_y$ layer coincide with regions exhibiting a high zigzag angle. This suggests an unusual in-plane Ni-O-Ni distortion due to F intercalation, a phenomenon not previously reported, a likely related to the two-step synthesis approach used in this work.

Figure 2f presents a magnified ADF image highlighting the pronounced zigzag arrangement of Nd atoms. In principle, the corresponding structural model would be expected to be the orthorhombic structure depicted in Figure 2g, where two neighboring oxygen atoms reside at each basal site from this viewing direction. However, due to the limitations of TEM resolution, these adjacent oxygen atoms are difficult to distinguish, resulting in a near-zero in-plane Ni-O-Ni angle in this projection. Remarkably, the magnified iCoM image in Figure 2h unveils an intriguing structural distortion where the basal oxygen simultaneously shifts in the same direction as the Nd atoms, as illustrated by the green polyhedral model in Figure 2i. Furthermore, the iCoM image in Figure S3 exhibits a comparable structural distortion consistent with Figure 2i, alongside a significant polyhedral rotation in a region lacking Nd displacement. These iCoM images definitively demonstrate the substantial differences in atomic arrangement between the two observed domains within the anion sublattice.

To definitively confirm F-ion intercalation within the NdNiO$_2$ layer, we employed high-resolution STEM-EELS to map the elemental distribution across the interface in the NdNiO$_2$/SrTiO$_3$ superlattice (Figure 3). The HAADF image in Figure 3a demarcates the region analyzed by EELS, where the distinct contrast readily differentiates the NdNiO$_2$ and SrTiO$_3$ layers. The corresponding elemental maps in Figure 3b, acquired at the Nd M$_{4,5}$, Ni L$_{2,3}$, Sr L$_{2,3}$, Ti L$_{2,3}$, O K, and F K edges, visualize the atomic-scale distribution of constituent elements. Notably, a prominent F signal is present within the NdNiO$_2$ layer and absent in the SrTiO$_3$ layer.

Figure 3c depicts the row-averaged line profile of the oxygen signal intensity extracted from the region marked in Figure 3a. The gradual decrease in oxygen intensity from the interface towards the inner nickelate layer reflects the variation in oxygen concentration. Conversely, the F-signal intensity in Figure 3d exhibits a gradual increase moving inwards from the interface. This observation suggests a higher concentration of F ions in the inner NdNiO$_2$F layer compared to the interfaces. This difference can be attributed to the presence of residual oxygen at the interfaces within our NdNiO$_2$/SrTiO$_3$ superlattice, as previously reported(*13*).

Furthermore, to estimate the oxidation state of Ni atoms in the NdNiO$_2$F layer, we quantified the Ni L$_3$/L$_2$ ratio (Figure 3e). The Ni valence state within the inner nickelate layer fluctuates around 2+, with occasional slight deviations below 2+. This finding suggests a possible composition of either NdNiO$_2$F or NdNiO$_2$F$_x$ (where *x* < 1). In contrast, the Ni valence state near the interfaces exhibits an intermediate state between $Ni^{2+}$ and $Ni^{3+}$. This observation can be attributed to the influence of residual oxygen at the interface*(36)*, potentially leading to an $NdNiO_{2.5}F_{0.5}$ configuration.

Theoretical calculations on various oxyfluorides predict anion ordering based on energetic stability. These calculations typically suggest F-ion intercalation at either apical or basal sites of the oxygen octahedra, or in interstitial sites within Ruddlesden-Popper structures*(4, 27-31)*. Given our two-step synthesis approach, F-ion intercalation within the apical sites was anticipated, leading to an ordered O/F configuration. To determine the specific distribution of F ions within the infinite layer structure, we meticulously quantified EELS measurements at targeted positions (Figure 4). The HAADF image in Figure 4a depicts the atomic structure chosen for EELS analysis. Red and yellow dots mark the targeted apical and basal positions within the polyhedra, respectively. To minimize the influence of signal mixing, EELS measurements were performed in a region with a t/λ value of ~0.2 (see Figure S4), corresponding to a sample thickness of approximately 15 nm. Since beam scattering at this thickness can introduce a degree of signal mixing from adjacent columns into the EELS results, we performed the atomically resolved EELS simulation shown in Figure S5. There is ~6% of the O signal at the F site and ~16% of the F signal at the O site, where the regions for calculation are marked in the composite map in Figure S5f. To determine the position of the intercalated F-ions, we extracted EELS spectra corresponding to the F-K edge at both the apical and basal positions, as shown in Figures 4b and 4c. Notably, both positions exhibit a significant F signal, with approximately the same F/O signal intensity ratio at both apical and basal positions in the inner nickelate layer (see Figure S6), indicating the F intercalation locally at both apical and basal sites.

To elucidate the F/O configuration and its impact on structural distortion, we employed DFT calculations in conjunction with analysis of the observed distortions. The HAADF image in Figure 4d reveals a significantly elliptical Nd column. This observation aligns well with the relaxed structural model in Figure 4e, which depicts F-ion intercalation at the apical sites. The elliptical shape of the Nd column arises from a viewing direction mismatch, leading to a more pronounced effect compared to perovskite nickelate (see Figure S7). An alternative structural model is presented in Figure 4f, featuring a 90° rotation compared to Figure 4e. This rotation results in a distinct atomic arrangement with a different zigzag distortion of the Nd atoms. Interestingly, in localized regions, we observe adjacent Nd columns with elliptical shapes oriented perpendicular to each other, as shown in Figure 4g. This experimental finding aligns with the DFT-calculated model in Figure 4h, which suggests F-ion intercalation at the basal sites. This observation further corroborates the EELS measurements presented earlier. A 90° rotation of the viewing direction in Figure 4i, relative to Figure 4h, reveals a similar zigzag distortion of the Nd atoms as in Figure 4f. However, a slightly stronger polyhedral distortion is evident, manifested by a greater separation of the adjacent anions at both apical and basal sites.

We now turn our attention to the anion sublattice within the F-intercalated $NdNiO_2$ layer. The iCoM image in Figure 5a depicts a local region exhibiting a typical orthorhombic structure with evident octahedral rotation. Additionally, a slight mismatch of the Nd columns is observed from this viewing direction. Notably, the intercalation of F ions at apical sites, as shown in the relaxed structure model under $SrTiO_3$ strain (Figure 5b), leads to an increased Nd mismatch, consistent with the elliptical Nd columns observed in the HAADF image (Figure 4d). A 90° rotation of the viewing direction (from [100] to [010]) in Figure 5c reveals a distinct atomic arrangement. Here, a pronounced zigzag displacement of Nd atoms is evident, while no visible polyhedral distortion is observed. This finding aligns well with the relaxed structure model presented in Figure 5d. Interestingly, the presence of two structural configurations within a single viewing direction (Figures 1e and S2) suggests the coexistence of widely distributed domains A and B in the F-intercalated $NdNiO_2$ layers. Domain A is the main configuration in the $NdNiO_xF_y$ layers. Domain B, with obvious Nd zigzag arrangement, is randomly distributed within a few nanometers. Figure 5e showcases a relaxed structure model depicting these two domains, highlighting the clear structural coupling at the domain interface marked by the yellow box.

The iCoM image in Figure 5f provides compelling experimental evidence for this structural coupling at the domain interface. Here, we observe competition between the geometry frustration-induced Nd ion displacement and the polyhedral distortions, mirroring the calculated relaxed structure in Figure 5g. Furthermore, the iCoM images in Figures 5h and 2h reveal an intriguing structural distortion where the basal anion displacement direction aligns with that of the Nd atoms. This observation points towards a configuration distinct from the F-ion intercalation at apical sites depicted in Figures 5c and 5d. As shown in Figure S8b, F intercalation at basal positions can lead to a slightly greater polyhedral distortion compared to intercalation at apical positions. This distortion can induce a shift in one of the basal F-atoms, bringing it closer to the Nd atoms. This phenomenon explains the weak residual contrast observed between Nd columns and basal F ions in Figures 2h and S8a (marked by red arrows). For regions exhibiting no visible residual contrast between Nd and basal anions (Figures 5h and S8c, marked by yellow arrows), we propose a composition of $NdNiO_2F_{0.5}$. This model incorporates partial F-atom intercalation within the $NdNiO_2$ layer, as shown in the relaxed structure model of Figure 5i. This configuration effectively explains the observed absence of residual contrast.

The interplay of anion intercalation and deintercalation in nickelates triggers significant structural modifications. In the reduced $8NdNiO_x/4SrTiO_3$ superlattice sample, residual oxygen persists at the interfaces and inner layers despite the stabilization of the infinite layer structure. This residual oxygen impacts F-ion incorporation within the reduced nickelate. Chemical composition measurements and Ni valence quantification suggest a configuration near the interfaces resembling $NdNiO_{2.5}F_{0.5}$ with a polyhedral $NiO_5F$ structure. The coupling between $TiO_6$ octahedra and $NiO_5F$ polyhedra induces subtle Nd atom displacements alongside polyhedral rotations within the interfacial nickelate layers.

In contrast, F-ion intercalation within the infinite layer structure leads to a substantial geometric reconstruction, resulting in an orthorhombic structure akin to perovskite nickelate, but with a more pronounced distortion. Due to the two-step fluorination process, F ions are primarily intercalated at the apical sites of the polyhedron, with partial intercalation at the basal sites. We have distinguished the F-ion intercalation positions (basal vs. apical) based on the varying cation displacements observed in HAADF images, which aligns with the EELS-determined chemical composition. Intriguingly, two widely distributed domains form within the nickelate layers. Near the domain interface region, a competition arises between the geometry frustration-induced Nd displacement and the polyhedral rotations. Additionally, a unique structural distortion is observed within domain B, potentially arising from the partial intercalation of F ions at basal sites and exhibiting an O/F ordered arrangement according to DFT calculations.

Furthermore, DFT calculations incorporating on-site Coulomb interaction (U) elucidate the influence of F-ion intercalation-induced distortions on the nickelate's electronic structure. F-ion intercalation drives the Ni valence state towards 2+, consequently inducing a distinct band gap(*1*), as reflected in the density of states for the $NdNiO_2F$ configuration (Figure S9). Resistance measurements of the fluorinated superlattice show highly insulating behavior, so that it was not possible to measure its temperature dependence. Notably, the valence band is dominated by hybridized Ni/O bands, with the F band positioned far from the Fermi level in the $NdNiO_2F$ structure. As shown in Figure S9 for various F-intercalation configurations, the band gap is primarily influenced by the overall F-ion concentration, with a minor dependence on the specific F-ion position (basal vs. apical). This can be attributed to the modification of electron occupancy in the Ni orbitals upon F-ion intercalation within the infinite layer structure. The on-site Coulomb interaction subsequently suppresses electron hopping between neighboring Ni sites in the $NdNiO_2F$ structure(*4*).

In summary, our findings demonstrate that the infinite layer structure serves as a compelling model system for tailoring anion intercalation. We present direct observations of the geometric reconstruction process triggered by F-ion intercalation within the nickelate's infinite layer structure. Due to the comparable ionic radii of $F^-$ and $O^{2-}$, F-ion incorporation leads to the formation of a sixfold-coordinated orthorhombic structure akin to perovskite nickelates, but exhibiting a more pronounced distortion. Notably, residual oxygen persisting at the interfaces in the reduced superlattice sample can promote the formation of a $NiO_5F$ configuration upon fluorination. Furthermore, we reveal the formation and widespread distribution of two distinct domains within the nickelate layers. These domains exhibit significant coupling at the domain wall, leading to a competition between the geometry frustration-induced Nd displacement and polyhedral distortion. We also observed F-ion intercalation at basal sites in localized regions, evidenced by the distinct elliptical direction of adjacent Nd columns. This observation is further corroborated by high-resolution EELS measurements. Additionally, a unique structural distortion was identified, characterized by the alignment of Nd atom and basal anion displacements. This phenomenon, supported by DFT calculations, suggests partial F-ion intercalation at basal sites, providing evidence for anion mobility during the fluorination process.

Our findings highlight the NdNiO$_2$F configuration as the dominant phase within the nickelate layer. Here, the on-site Coulomb repulsion between neighboring Ni$^{2+}$ cations effectively suppresses electron transport. Since F-ion intercalation primarily influences the Ni electronic configuration, precise control over this process within the infinite layer nickelate system appears critical for achieving the desired Ni 3d$^{8.8}$ configuration, a potential route towards superconductivity(*3, 32*). Furthermore, previous reports suggest that the electronic structure modifications induced by F-ion incorporation in nickelates can be reversed by subsequent oxygen intercalation through annealing in an oxygen atmosphere(*1*). This opens avenues for the development of resistance switching materials. Our work also sheds light on the strategic application of combined anion intercalation and deintercalation for investigating the ordered arrangement of anions in related systems. The high-spin state of Ni$^{2+}$ in the F-intercalated infinite layer structure underscores the significant influence of ordered/disordered anion arrangements on the magnetic structure of nickelates, as previously reported(*27, 29*). This dependence highlights the intricate interplay between anion distribution and magnetic properties, paving the way for the tailored manipulation of magnetic behavior through controlled manipulation of the anion arrangement. In conclusion, elucidating the effects of F-ion intercalation on the atomic and electronic structures within the infinite layer structure system holds broader implications for perovskite oxyfluoride materials, where both polyhedral distortion and anion order/disorder play prominent roles.

**Methods**

**Materials and sample preparation.** The NdNiO$_3$/SrTiO$_3$ superlattice samples were epitaxially grown on a (001)-oriented single crystal SrTiO$_3$ substrate by pulsed laser deposition (PLD) technique according to the procedures described in ref.(*13, 24, 33*). Subsequently, the topotactical reduction was performed with CaH$_2$ powder inside a vacuum sealed Pyrex glass tube at a temperature of 280°C for aproximately four and a half days. The preparetion of the sample with the CaH$_2$ powder and the glass tube was performed inside an Ar-filled glove box to prevent any oxygen contaminants. (*3*), (*13, 33*) The fluorination process is then performed using polyvinylidene fluoride (PVDF). *(27, 31)* The process took place in a tube furnace which was filled with a constant Ar flow before and during the chemical reaction. The heater was ramped up at approximately 5°C/min until the furnace reached a stable temperature of 350°C. The process was then continued for one hour and the heater was turned off. The system was allowed to cool down to room temperature, which usually took between 4 and 5 hours. After this, the sample was removed and cleaned with Acetanol. The TEM lamellae were prepared using a focused ion beam (FIB Scios, FEI) in a high-vacuum environment.(*34*) To mitigate beam-induced charging effects during sample preparation, a thin layer of approximately 6 nm of carbon was deposited on the sample surface using a high-vacuum sputter coater (EM ACE 600, Leica). A Fischione NanoMill® TEM sample preparation system was used to improve the quality of the TEM lamellae through low-energy milling and high-vacuum cleaning.

**STEM imaging and STEM-EELS simulation.** The STEM studies were performed using a JEOL JEM-ARM200F microscope (JEOL Co. Ltd.) equipped with a DCOR probe corrector and a Gatan GIF Quantum ERS K2 spectrometer. For STEM imaging, a condenser aperture of 30 μm was used, with a corresponding convergence semi-angle of 20.4 mrad. We use a collection semi-angle between 83 and 205 mrad for HAADF imaging, and 85 mrad for EELS measurements. A dispersion of 0.5 eV/channel with an energy resolution of about 1 eV was used to acquire the EELS spectra. The 4D-STEM data set was acquired using a Merlin pixelated detector (256 × 256 pixels, Quantum Detectors) in 1-bit mode, with continuous read/write at a pixel time of 48 μs. Post-processing of the 4D-STEM dataset is based on the Python libraries of py4dstem(*35*) and fpd(*36*). The atomically resolved STEM-EELS simulations were performed using an open-source Python package from abTEM.(*37*) A 4 × 4 × 40 unit cell superstructure, corresponding to the same sample thickness as we studied experimentally, was used for the calculations. The atomic potential was taken from the independent atom model potential.(*37*) The beam energy, convergence angle, and collection angle in the simulations are the same as those used in the experiment.

**DFT calculations.** First-principles DFT calculations were performed to investigate the structure distortion and the density of states (DOS) variation of the F-ion intercalation in the infinite layer nickelate in the $NdNiO_2/SrTiO_3$ superlattice. The calculations used the Generalized Gradient Approximation (GGA) and the Perdew-Burke-Ernzerhof (PBE) functional for exchange correlation, implemented in the Vienna Ab initio Simulation Package (VASP).(*38, 39*) The plane-wave cutoff energy of 520 eV was used. The DFT+U method used a Hubbard U parameter of 4.0 eV for the Ni atoms. For the structure optimization, the maximum ionic force is set to 0.01 eV/Å, the self-consistent convergence of the total energy is $5 \times 10^{-7}$ eV/atom and the maximum ionic displacement is set to 0.5 Å. The VASP calculated data were analyzed using the VASPKIT code(*40*).

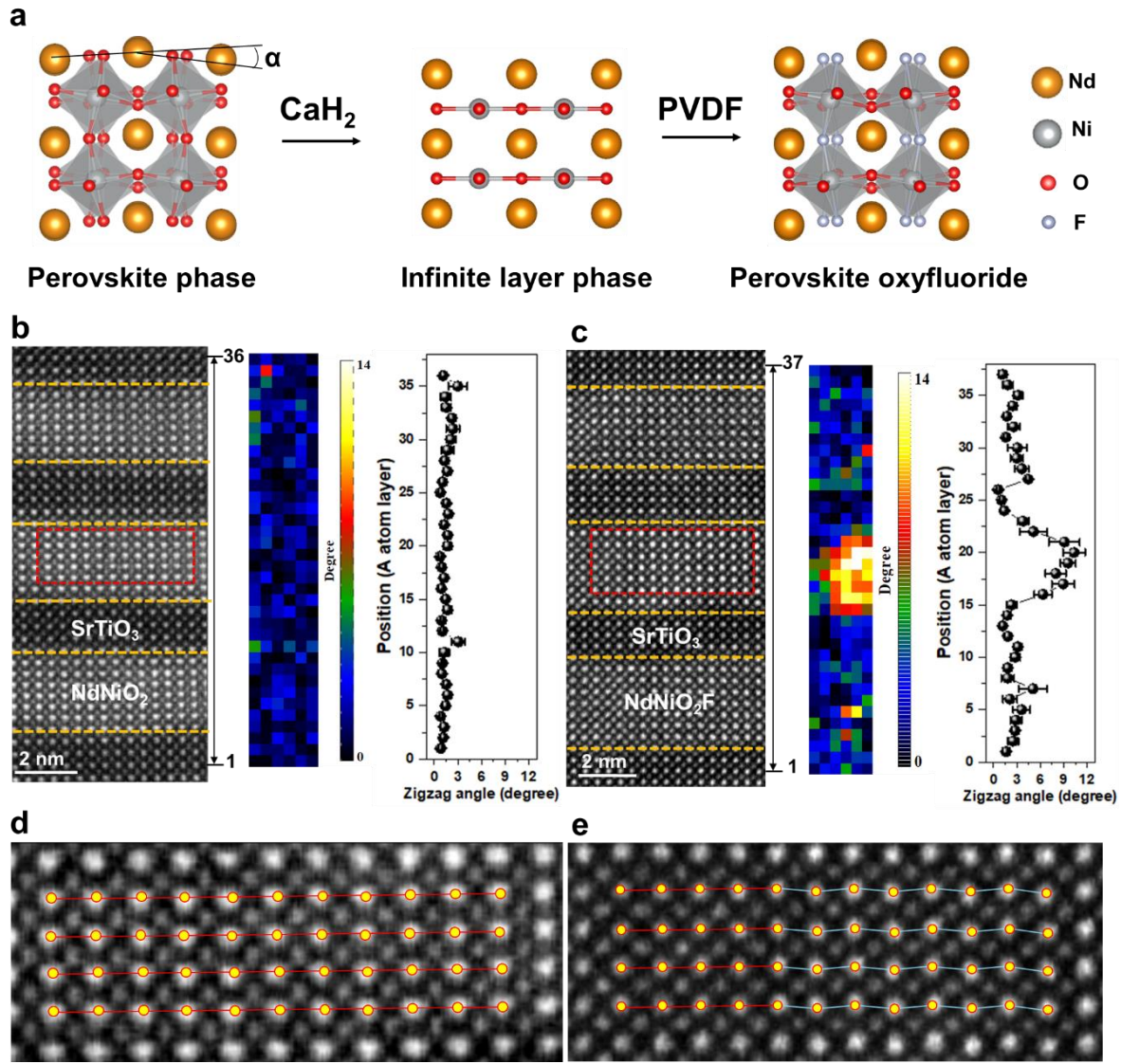

**Figure 1:** Synthesis of the fluorine-intercalated infinite layer nickelates and the formation of different domains in the oxyfluoride NdNiO$_2$F. (a) Structural models of the perovskite phase, the infinite layer phase and the oxyfluoride phase. (b) HAADF image of the 8NdNiO$_2$/4SrTiO$_3$ superlattice, a map of the zigzag angle α, and a line profile of the zigzag angle α between the A-site cations (A: Nd and Sr). (c) HAADF image of the 8NdNiO$_2$F/4SrTiO$_3$ superlattice, a map of the zigzag angle α, and a line profile of the zigzag angle α between the A-site cations (A: Nd and Sr). (d) A magnified HAADF image of the region marked by the red dashed box in (b). (e) A magnified HAADF image of the region marked by the red dashed box in (c) showing two domains. In (d) and (e), the yellow dots correspond to the A-site cation positions determined by Gaussian fitting, and the red and blue lines are guides for the eye showing straight and zigzag cation arrangements, respectively.

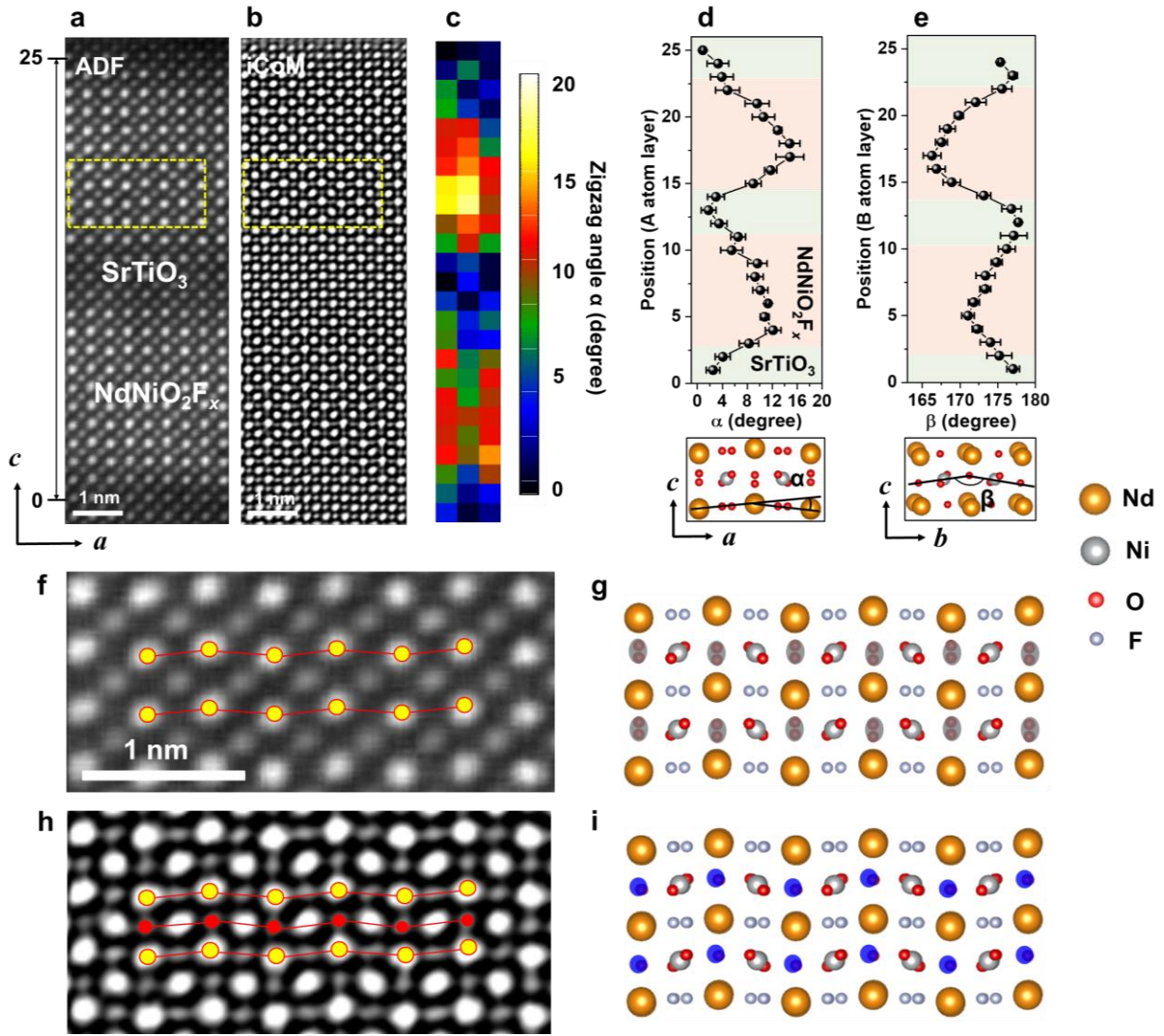

**Figure 2:** Atomic distortion in the oxyfluoride $NdNiO_2F_x$. (a) ADF and (b) iCoM images of the 8$NdNiO_2$F/4$SrTiO_3$ superlattice reconstructed from the 4D-STEM data set. Quantified zigzag angle α (c) map and (d) line profile of the A-site cations (A: Nd, and Sr) in the region marked from the 1st to the 25th unit cell in (a). (e) B-C-B (B: Ni, and Ti, C: basal anions) angle plot of the same region in (d). The green and red shaded areas in (d) and (e) correspond to $SrTiO_3$ and $NdNiO_2F_x$, respectively. (f) Enlarged ADF image of the region marked by the yellow dashed box in (a) and the estimated structure model in (g). (h) Enlarged iCoM image of the region marked by the yellow dashed box in (b) and the estimated structure model in (i).

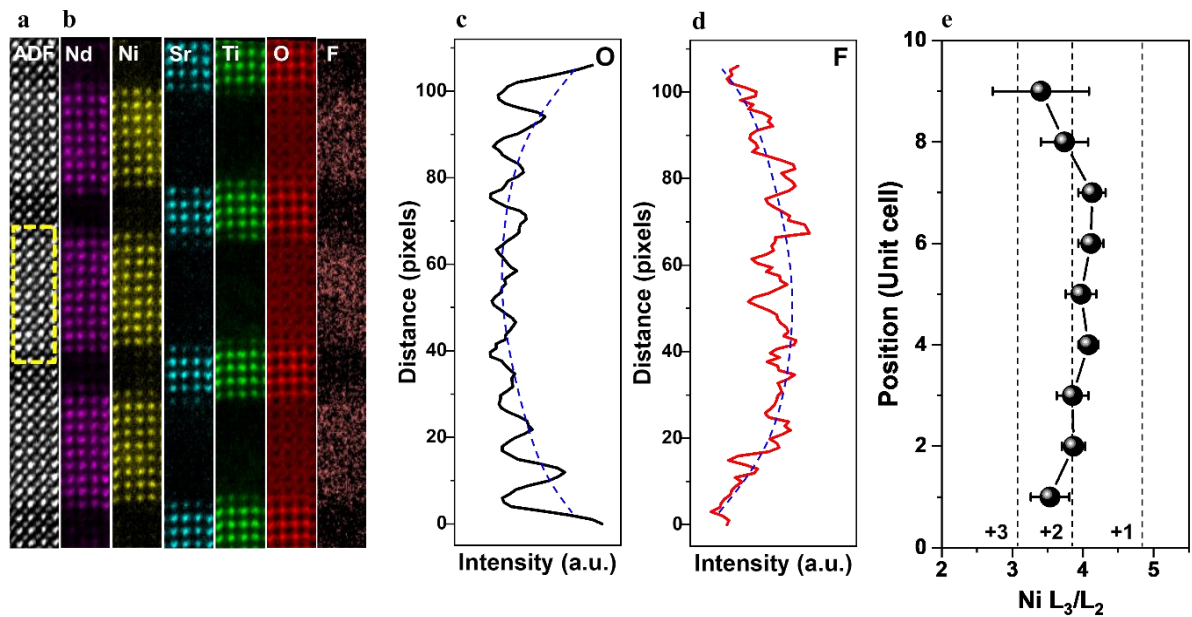

**Figure 3:** Elemental distribution, composition, and valence state in the oxyfluoride $NdNiO_2F$ superlattice. (a) ADF image for the EELS measurement of the $8NdNiO_2F/4SrTiO_3$ superlattice. (b) The corresponding element distributions: Nd (purple), Ni (yellow), Sr (blue), Ti (green), O (red) and F (orange). (c) Intensity line profile of the O K edge signal extracted from the region marked by the yellow dashed box in (a). (d) Intensity line profile of the F K edge signal extracted from the region marked by the red dashed box in (a). (g) The calculated Ni $L_3/L_2$ intensity ratio in the nickelate layer. The dashed lines indicate the reference values of the Ni $L_3/L_2$ intensity ratio for different Ni valence states, which are derived from $NdNiO_3$ for $Ni^{3+}$,(*41*) NiO for $Ni^{2+}$,(*42*) and $NdNiO_2$ for $Ni^+$.(*13*)

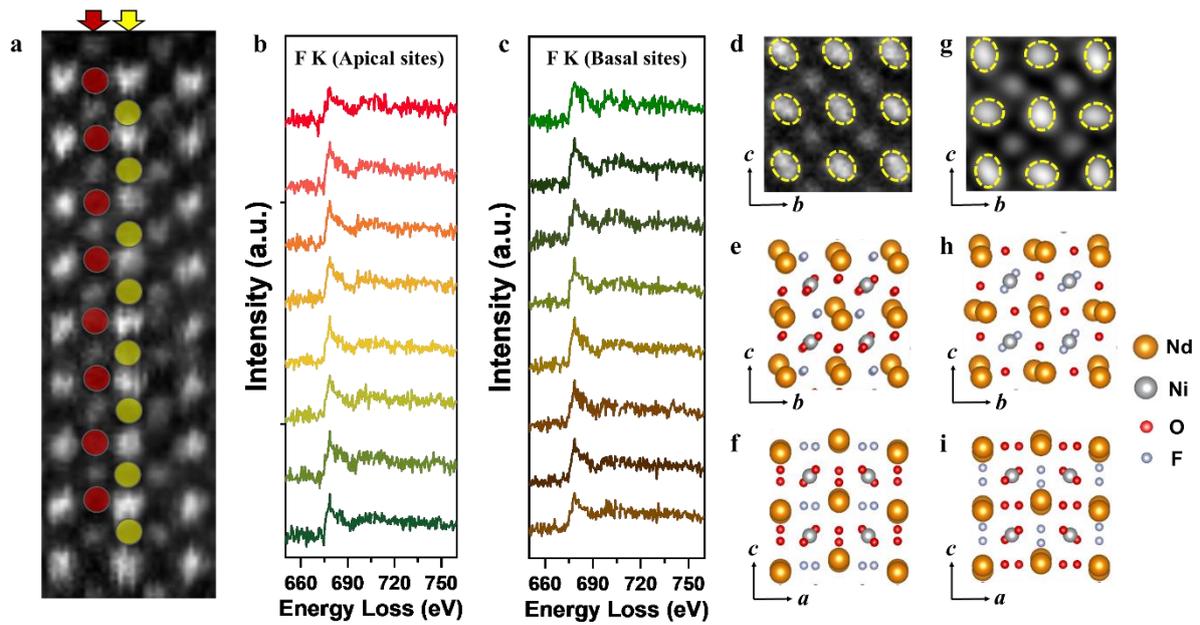

**Figure 4:** Detected F-signal at both basal and apical sites and its effect on the atomic structure distortion. (a) HAADF image of the region for EELS measurements. Red and yellow spots mark the apical and basal sites, respectively. (b) The F signal extracted from the red spots in (a). (c) The F signal extracted from the yellow spots in (a). (d) HAADF image of the region with elliptical Nd columns pointing in the same direction. The corresponding structure models with F intercalation at apical sites for viewing directions [100] in (e) and [010] in (f), respectively. (g) HAADF image of the region with perpendicular elliptical Nd columns. The corresponding structure models with F intercalation at basal sites for the viewing directions (h) [100] and (i) [010], respectively.

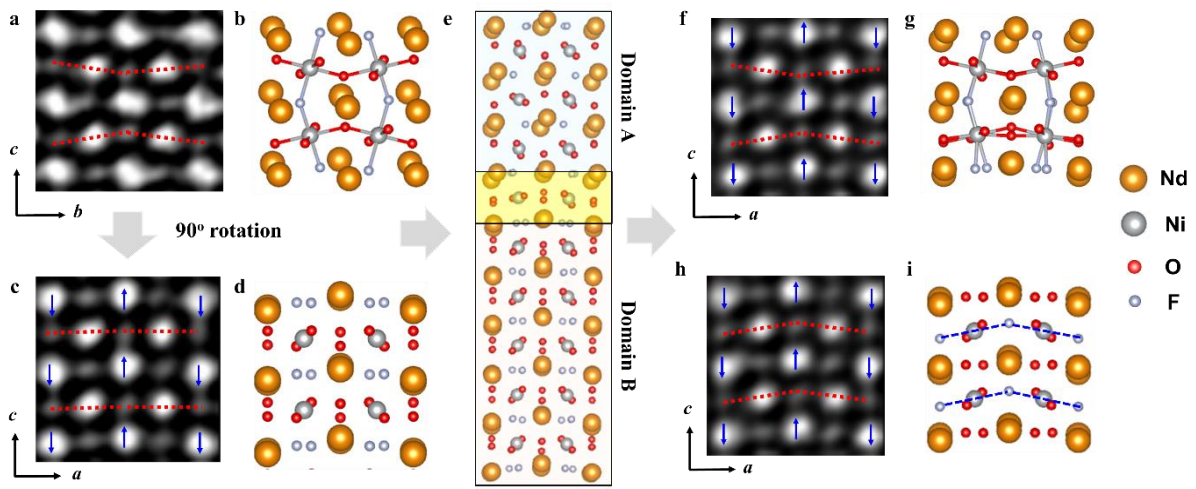

**Figure 5:** Comparison between the experimentally observed structure distortion and the relaxed structure model from DFT calculations. (a) Reconstructed iCoM image of domain A with clear polyhedral rotation and (b) the relaxed structure model with F intercalation at apical sites. The red dashed lines mark the anion sublattice distortion. (c) Reconstructed iCoM image of domain B showing a clear zigzag arrangement of Nd atoms and (d) the relaxed structure model with F intercalation at apical sites. The blue arrows indicate the displacement direction of the Nd atoms. The red dashed line marks the anion sublattice without distortion. Domain B is formed by a 90° rotation of domain A. (e) A relaxed structural model with two connected domains. (f) Reconstructed iCoM image of the domain interface and the corresponding enlarged structural model. (h) Reconstructed iCoM image of the region with abnormal structure distortion and (i) the corresponding structure model from the DFT calculations with partial intercalation of F ions at the basal sites.

**Data availability**

The data that support the findings of this study are available from the corresponding author on reasonable request.

**Acknowledgments**

This project has received funding from the European Union's Horizon 2020 research and innovation programme under Grant Agreement No. 823717-ESTEEM3. The authors are thankful to Dr. Y. Wang for the support of 4D-STEM data processing, Dr. T. Heil for the support of merlin software, Dr. K. Kern, Dr. J. Tatchen, Dr. A. Schuhmacher, and X. J. Zhang for the support of the supercomputer software, K. Hahn and P. Kopold for TEM support, and Dr. D. S. Weng for the discussion of DFT calculations. The authors would like to thank Roland Eger for his help in setting up the fluorination setup.

**Author Contributions Statement**

C.Y. and R.A.O. conceived the project. C.Y., H.G.W., W. S., and P.A.v.A. conducted the STEM measurements and related data analysis. R.A.O. grew the samples and performed the totapical reduction and fluorination. P.A.v.A, E.B, and B.K. supervised this work. C.Y. did the simulations and calculations. K.A. provided insight and discussion of the DFT results. C.Y. wrote the paper with contributions from all authors. All authors contributed to the discussion and comments.

**Competing Interests Statement**

The authors declare no competing financial or non-financial interests.